\documentclass[conference]{IEEEtran}
\IEEEoverridecommandlockouts
\usepackage{tcolorbox}
\usepackage{cite}
\usepackage{amsmath,amssymb,amsfonts}
\usepackage{algorithmic}
\usepackage{graphicx}
\usepackage{textcomp}
\usepackage{xcolor}
\usepackage{times}
\usepackage{latexsym}
\usepackage{hyperref}
\usepackage{multirow}
\usepackage{url}
\usepackage{balance}
\usepackage{bm}
\usepackage{amssymb}
\usepackage{amsmath}
\usepackage{url}
\usepackage{makecell}
\usepackage{multirow}
\usepackage{color}
\usepackage{colortbl}
\usepackage{CJK}
\usepackage{enumitem}
\usepackage{booktabs}
\usepackage{microtype}
\usepackage{arydshln}
\usepackage{amsfonts}
\usepackage{eufrak}
\usepackage{balance}
\usepackage{listings}
\usepackage[T1]{fontenc}

\usepackage[utf8]{inputenc}

\usepackage{microtype}
\usepackage{pifont}
\usepackage{inconsolata}
\def\BibTeX{{\rm B\kern-.05em{\sc i\kern-.025em b}\kern-.08em
    T\kern-.1667em\lower.7ex\hbox{E}\kern-.125emX}}

\definecolor{codegreen}{rgb}{0,0.6,0}
\definecolor{codered}{rgb}{0.6,0,0}
\definecolor{codegrey}{rgb}{0.5,0.5,0.5}

\lstdefinestyle{diff}{
    morecomment=[f][\color{codegreen}]{+},
    morecomment=[f][\color{codered}]{-},
    morecomment=[f][\color{codegrey}]{@@},
    basicstyle=\ttfamily\small,
    breaklines=true,
    captionpos=b
}

\begin{document}

\title{Using Large Language Models for Commit Message Generation: A Preliminary Study}

\author{\IEEEauthorblockN{Linghao Zhang, Jingshu Zhao, Chong Wang$^{*}$, Peng Liang$^{*}$}
\IEEEauthorblockA{
School of Computer Science, Wuhan University, Wuhan, China\\
\{starryzhang, candyzhao, cwang, liangp\}@whu.edu.cn}
}

\maketitle

\begin{abstract}
A commit message is a textual description of the code changes in a commit, which is a key part of the Git version control system (VCS). It captures the essence of software updating. Therefore, it can help developers understand code evolution and facilitate efficient collaboration between developers. However, it is time-consuming and labor-intensive to write good and valuable commit messages. Some researchers have conducted extensive studies on the automatic generation of commit messages and proposed several methods for this purpose, such as generation-based and retrieval-based models. However, seldom studies explored whether large language models (LLMs) can be used to generate commit messages automatically and effectively. To this end, this paper designed and conducted a series of experiments to comprehensively evaluate the performance of popular open-source and closed-source LLMs, i.e., Llama 2 and ChatGPT, in commit message generation. The results indicate that considering the BLEU and Rouge-L metrics, LLMs surpass the existing methods in certain indicators but lag behind in others. After human evaluations, however, LLMs show a distinct advantage over all these existing methods. Especially, in 78\% of the 366 samples, the commit messages generated by LLMs were evaluated by humans as the best. This work not only reveals the promising potential of using LLMs to generate commit messages, but also explores the limitations of commonly used metrics in evaluating the quality of auto-generated commit messages. 



\end{abstract}


\section{Introduction}

A commit is an operation to submit updates to the version control system (VCS), which is the key operation of Git VCS. Each commit should be accompanied by a commit message, which serves as a concise textual description of the update. Commit messages can record changes of the source code, help developers understand the code, and facilitate efficient collaboration. However, manually writing commit messages is time-consuming, and the commit messages sometimes are non-informative or even empty\cite{change}. To address this problem, researchers have proposed techniques for automatically generating commit messages which achieved some success\cite{evaluation}.

Large Language Model (LLM) is a popular artificial intelligence technology that uses deep learning and massively large data sets to understand, summarize, generate, and predict new text content. LLMs often have an extremely large number of parameters and are trained on huge datasets. Nowadays, the application of LLMs in the field of software engineering has achieved great success\cite{hou2023large} and attracted the attention of researchers. Recent advances in LLMs are very promising as reflected in their capability for general problem-solving in few-shot and zero-shot setups, even without explicit training on certain tasks. However, there is still a lack of attention on the ability of LLMs to automatically generate commit messages. 

In this work, we conducted a preliminary evaluation to explore the feasibility and performance of LLMs on commit message generation task. In addition to the traditional metrics BLEU~\cite{bleu} and Rouge-L~\cite{rouge}, we also conducted the human-ChatGPT evaluation. The results show that LLMs have comparable performance to existing methods on BLEU and Rouge-L metrics. In the human evaluation, surprisingly, LLMs beat other methods by an absolute margin. In 78\% of 366 samples, the commit message generated by LLMs was considered the best by human participants. Interestingly, we also included human-written commit messages as one of the options, and only 13.1\% of the 366 samples were manually evaluated as the best. This work highlights the potential of LLMs in commit message generation task and underscores the limitations in using existing metrics to evaluate the quality of commit messages.

The main contributions of this work are as follows:
\begin{itemize}
    \item To provide a 2-phase evaluation study to explore the feasibility and performance of LLMs on commit message generation.
    \item To explore the weakness of BLEU and Rouge-L metrics in evaluating the quality of auto-generated commit messages.
\end{itemize}

The rest of this paper is organized as follows. Section~\ref{sec:related} provides existent studies on commit message generation and applications of LLMs. Section~\ref{sec:design} presents the research questions of this work and the overview of our approach. Section~\ref{sec:result} is the results and discussion.

The evaluation data and scripts used in this work have been made available online~\cite{data}.

\begin{figure*}[!t]
\centering
\includegraphics[width=\textwidth]{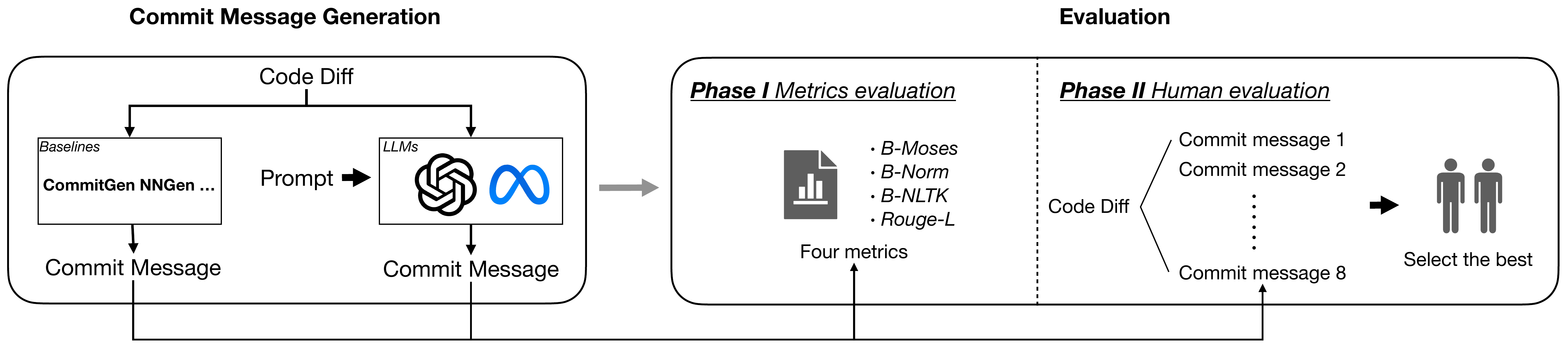}
\caption{Overview of our approach}
\label{fig:overview}
\end{figure*}

\section{Related Work}
\label{sec:related}
In recent years, some researchers explored the automatic generation of commit messages. According to the classification of the commit message generation models in~\cite{evaluation}, the proposed methods can be grouped into the following two categories.

\textbf{Generation-based methods} normally utilize the encoder-decoder architectural framework and are trained on the datasets collected from open-source repositories. For example, CommitGen~\cite{commitgen} is an early attempt to use neural machine translation in commit message generation. NMT~\cite{nmt} is another neural machine translation method similar to CommitGen, but uses a different attention mechanism. CoDiSum~\cite{codisum} uses a multi-layer bidirectional GRU (Gated Recurrent Unit) as its encoder part, which can better learn the representations of code changes. PtrGNCMsg~\cite{ptrgncmsg} is an attentional RNN encoder-decoder model to translate code diffs to commit messages.

\textbf{Retrieval-based methods} retrieved the most relevant item in the dataset for each request. For instance, NNGen\cite{nngen} leverages the nearest neighbor (NN) algorithm to generate commit messages. To generate a commit message, NNGen calculates the cosine similarity between the target code diff and each code diff in the training set. Then, the top-k code diffs in the training set are selected to compute the BLEU scores between each of them and the target code diff. The one with the largest BLEU score is regarded as the most similar code diff, and its commit message will be used as the target commit message.

These existing studies revealed a rich landscape of methodologies aiming at automatic generation of commit messages. Our work stands apart from these existing methods by introducing the novel application of LLMs, specifically ChatGPT and Llama 2, to this task. The methods introduced in this section will be used as the baselines for our evaluation and comparative analysis.






\section{Research Design}
\label{sec:design}

\subsection{Research Question}

The research objective of this work is to explore the feasibility and effectiveness of LLMs in commit message generation. For this purpose, we define one research question (RQ): \textit{Can LLMs generate commit messages efficiently and accurately?} This RQ is designed to explore the feasibility of LLMs to automatically generate commit messages and to evaluate the performance of LLMs in this task. The answer to this RQ would help us understand whether LLMs are capable of efficiently and accurately generating commit messages. 


\subsection{Overview of Our Approach}

Fig. \ref{fig:overview} shows the overview of our approach. First, we perform the task of commit message generation with two LLMs, i.e., ChatGPT and Llama 2. Then, a 2-phase evaluation process is designed to evaluate the quality of generated commit messages by LLMs. Specifically, four metrics (detailed in Section~\ref{sec:Phase1}) are employed in Phase I, since BLEU~\cite{bleu} and Rouge-L~\cite{rouge} metrics are widely used to evaluate the quality of generated commit messages in previous studies, such as~\cite{codisum}\cite{2017jiang}. In this phase, the metrics scores are used to compare the commit messages generated by various models with the original ones written by developers. In Phase II, human beings are employed to investigate that the commit message generated by which method fits best the corresponding code difference, as detailed in Section~\ref{sec:Phase2}.


\subsection{Dataset}
In this exploratory study, we reused the dataset employed in~\cite{codisum}\cite{2017jiang}. The dataset contains 7,661 pairs of code diff and its corresponding human-written commit messages, and the length of each included commit message is more than three words. These 7661 pairs were randomly selected from 90,661 code diff-commit message pairs, which were collected from the top 1,000 popular Java projects on GitHub. The dataset used for the study has been provided online~\cite{data} for replication.

\subsection{Selection and Settings of LLMs}
\label{sec:2LLMs}
In this work, we selected two representative LLMs to explore the performance of LLMs in commit message generation. More specifically, as a closed-source LLM proposed by OpenAI, \textbf{ChatGPT\footnote{\href{https://chat.openai.com/}{https://chat.openai.com/}}} is a ChatBot. Whereas, \textbf{Llama 2\footnote{\href{https://ai.meta.com/llama/}{https://ai.meta.com/llama/}}} is provided by Meta as a collection of foundation language models ranging from 7B to 70B parameters. Llama 2 is open-sourced, and all the variants are released in the research community. It marks a huge step forward for open-source LLMs. 

For ChatGPT, we employed the gpt-3.5-turbo model available in the OpenAI API\footnote{\href{https://platform.openai.com/}{https://platform.openai.com/}}. Its temperature value is between 0-1, and we selected a moderate temperature value of 0.5, with all other parameters left as default. For Llama 2, two chat versions of the model, i.e., Llama 2 7B and Llama 2 70B.\, are employed. The former is with 7 billion (7B) parameters, and the latter is with 70 billion (70B). To be consistent with ChatGPT, the temperature value of Llama 2 was also set to 0.5.

Due to the generality of LLMs, they can be adapted for downstream tasks simply by crafting an appropriate prompt. This method of directly applying to sub-domains without any fine-tuning is one of the distinct characteristics of LLMs, called `zero-shot prompting'~\cite{radford2019language}.
For either ChatGPT or Llama 2 variants, we manually constructed the following basic prompt to enable LLMs to generate commit messages based on code differences.

\texttt{The following is a diff which describes the code changes in a commit, Your task is to write a short commit message accordingly.}
\texttt{ [DIFF] }
\texttt{According to the diff, the commit message should be: } 

where the \texttt{[DIFF]} is code diff slot.

\subsection{Metrics and Baselines used in Evaluation: Phase I}
\label{sec:Phase1}


In machine translation and text summarization, BLEU~\cite{bleu} and Rouge-L~\cite{rouge} stand as the universal evaluation metrics to measure the quality of the generated text, by calculating the similarity score between auto-generated text and human-written text. The higher the score, the higher the quality. We employed the following three variants of BLEU~\cite{evaluation} and Rouge-L as the evaluation metrics in Phase I of our designed evaluation process.

\begin{itemize}
    \item \textbf{B-Moses} metric script comes from a well-known open-source toolkit
    , which is a widely used variant of the BLEU metric without a smoothing function. 
    \item \textbf{B-Norm} is a BLEU variant adapted from B-Moses. It converts all the characters to lowercase and adopts a smoothing method to smooth the n-gram precision scores. 
    \item \textbf{B-NLTK} is a BLEU variant metric from NLTK toolkit. It adopts a smoothing function that different from B-Norm.
    \item \textbf{Rouge-L} is a metric used in natural language processing, particularly for evaluating text summaries. It measures the longest common subsequence (LCS) between the generated summary and a reference summary. It is also widely used to evaluate the quality of generated commit messages.
\end{itemize}

In addition, the five methods proposed for commit message generation and introduced in Section~\ref{sec:related} will be used as the \textbf{baselines} in this preliminary study to be evaluated and compared with the LLMs.

\begin{figure}[b]
\centerline{\includegraphics[width=0.5\textwidth]{Figures/diff.pdf}}
\caption{Example of human evaluation}
\label{fig:diff}
\end{figure}

\subsection{Human Evaluation: Phase II}
\label{sec:Phase2}

To explore whether the auto-generated commit messages describe the differences between codes correctly, we designed and conducted a human evaluation on both the auto-generated commit messages and the ones provided by human developers. 


Based on statistical principles on representative samples in a dataset~\cite{israel1992dss}, we selected a size of 366 from the 7,661 samples based on 5\% margin of error and 95\% confidence level. These 366 code differences and their corresponding commit messages, which are auto-generated and written by human developers, are randomly selected to construct the dataset to be evaluated by human beings, as listed below. 

\begin{itemize}
    \item The code difference in the \textsc{.diff} files.
    \item Commit message generated by generation-based and retrieval-based methods mentioned in Section~\ref{sec:related}. 
    \item Commit message generated by the selected LLMs, as introduced in Section~\ref{sec:2LLMs}.
    \item Commit message written by human developers.
\end{itemize}

Fig.~\ref{fig:diff} shows an example data item in this dataset. Note that in the process of human evaluation, the order of candidate commit messages is shuffled. Then, the first and second authors independently reviewed each code difference and selected only one commit message, which fit the code difference best, out of the 8 candidates. Participants did not know the correspondence between commit messages and the methods that generate them respectively, which reduced subjective bias.


\section{Results \& Discussion}
\label{sec:result}


\subsection{Result of Phase I: Metrics evaluation}
TABLE~\ref{tab:BLEU} shows the scores of four metrics on the commit messages auto-generated by different models, including LLMs. It is observed that LLMs can achieve decent metrics scores compared with the five baseline models, as listed in the first column of TABLE \ref{tab:BLEU}. Regarding the B-Moses metric, ChatGPT ranks second, lagging behind the NNGen method, while Llama 2's score is lower. Considering the B-Norm metric, ChatGPT exceeds the average score of existing methods, while the scores of Llama 2 are not impressive. As for the Rouge-L metric, ChatGPT ranks second, lagging behind the CoDiSum. Particularly, ChatGPT got the highest score when considering the B-NLTK metric. In this case, Llama 2 70B outperformed Llama 2 7B counterpart but fell slightly behind CoDiSum.

\begin{table}[b]
\caption{Metric Scores on the Auto-generated Commit Message}
\begin{center}
\scalebox{1.1}{
\begin{tabular}{lcccc}
\toprule
Method & B-Moses & B-Norm & B-NLTK & Rouge-L\\
\midrule  \midrule
CommitGen\cite{commitgen} & 1.29 & 9.25 & 4.14 & 0.096 \\
CoDiSum\cite{codisum} & \cellcolor{gray!8}1.74 & \cellcolor{gray!25}\textbf{15.45} & \cellcolor{gray!8}5.72 & \cellcolor{gray!25}\textbf{0.187}\\
NMT\cite{nmt} & 1.24 & 9.75 & 3.74 & 0.100 \\
PtrGNCMsg\cite{ptrgncmsg} & 0.81 & \cellcolor{gray!16}12.65 & 4.79 & \cellcolor{gray!8}0.149\\
NNGen\cite{nngen} & \cellcolor{gray!25}\textbf{2.93} & 8.91 & 5.17 & 0.098\\
\midrule
ChatGPT & \cellcolor{gray!16}2.91 & \cellcolor{gray!8}11.84 & \cellcolor{gray!25}\textbf{7.39} & \cellcolor{gray!16}0.165\\
Llama 2 7B & 1.33 & 8.10 & 5.71 & 0.107\\
Llama 2 70B & 1.66 & 8.68 & \cellcolor{gray!16}7.08 & 0.135\\
\bottomrule
\end{tabular}
}
\label{tab:BLEU}
\end{center}
\end{table}

The results reveal that without fine-tuning or additional training, LLMs can yield decent results by constructing a basic prompt. Due to the easy use of LLMs, they can be seamlessly integrated into the integrated development environment (IDE), which further demonstrates the feasibility of LLMs.

\begin{tcolorbox}[sharp corners,boxrule=0.5pt]
\noindent
\textbf{Key findings of Phase I:} LLMs can be used for the automatic generation of commit message, and contribute decent performance in this task when evaluated with BLEU and Rouge-L metrics.
\end{tcolorbox}

\begin{figure}[b]
\centerline{\includegraphics[width=0.45\textwidth]{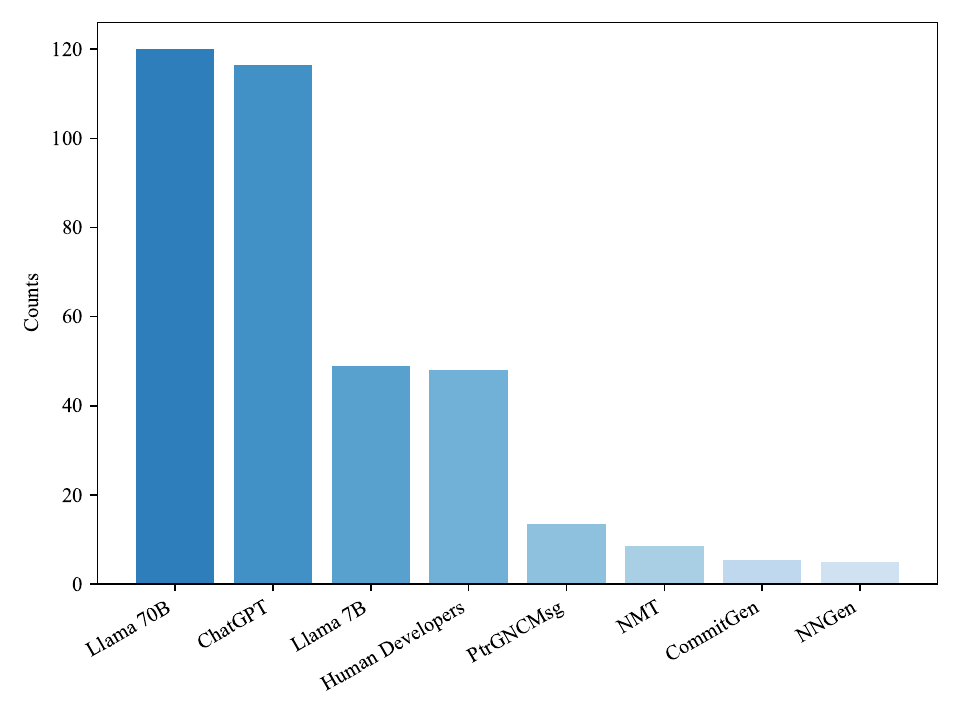}}
\caption{Human evaluation results (number of times selected as the best commit message corresponding to the code diff - average value of two participants)}
\label{fig:human}
\end{figure}

\begin{table}[h]
\caption{B-Norm score of each commit message in Fig. \ref{fig:diff}}
\begin{center}
\begin{tabular}{l p{5cm} c}
\toprule
\textbf{Method} & \textbf{Generated Commit Message} & \textbf{B-Norm}\\
\midrule  \midrule
ChatGPT & \textit{Set default value for name in MongoDBStorageProvider.} & 16.51 \\
\midrule
Llama 70B & \textit{Set default MongoDB storage provider name to `default'}. & 21.66 \\
\midrule
Llama 7B & \textit{Fixed naming convention for MongoDB storage provider class.} & 19.07 \\
\midrule
CommitGen & \textit{Modify reference to static field name.} & 22.83 \\
\midrule
PtrGNCMsg & \textit{Make default default `` default ''} & 0 \\
\midrule
NMT & \textit{Fix default stream name} & 21.41 \\
\midrule
NNGen & \textit{Initialize ExternalComponent with empty strings.} & 19.30 \\
\bottomrule
\multicolumn{3}{p{8cm}}{\textit{In this example, the commit message generated by ChatGPT is considered the best by both participants.} }
\end{tabular}
\label{tab:score}
\end{center}
\end{table}

\subsection{Result of Phase II: Human evaluation}
Following the human evaluation process designed in Section~\ref{sec:Phase2}, we got the best commit message suggested by the two coders, out of the 8 candidates. Fig.~\ref{fig:human} shows the distribution of their preferences on these 8 candidate commit messages after evaluating the 366 samples. Note that the commit messages generated by CoDiSum are omitted here, since we could not find its model file. We found that in 78\% of all the 366 samples, the commit messages generated by LLMs were considered the best among the 8 candidates. Specifically, 32.8\% and 31.8\% of 366 commit messages generated by Llama 2 70B and ChatGPT respectively were selected as the one to fit the corresponding code difference best.  
Considering the example in Fig. \ref{fig:diff}, for instance, the commit messages generated by LLMs are the most consistent with the code difference. On the contrary, the commit messages generated by the other methods cannot describe the meaning of code difference correctly or even be wrong. As TABLE~\ref{tab:score} shows, the commit message generated by ChatGPT got a lower score under the B-Norm metric, but was evaluated by human beings as the best commit message corresponding to the code difference. Although the baseline methods got high scores on the metrics, they are hardly favored by human participants. Interestingly, we also included human-written commit messages as an option, and commit messages generated by LLMs are evaluated as the best even more often than human-written ones.

To evaluate whether a method of generating commit messages is applicable or not, human evaluation results are the most direct indicator, since all methods of commit message generation are ultimately designed to serve developers. Based on the results of human evaluation, the performance of LLMs is better than others.

\begin{tcolorbox}[sharp corners,boxrule=0.5pt]
\noindent
\textbf{Key findings of Phase II:} LLMs demonstrate superior performance in generating commit messages, and they are more preferred by human developers compared to the existing methods.
\end{tcolorbox}

\subsection{Discussion}

\textbf{Our results further reveal the quality problems of human-written commit messages.} Loyola et al.~\cite{nmt} reported that 16\% of messages in a widely-used dataset were noisy, offering little useful information about the commit. Fig.~\ref{fig:human} shows that only 48 human-written commit messages, accounting for 13.1\% of the 366 samples, were selected as the best by the two coders.

\textbf{The quality of human-written commit messages may affect the applicability of the existing metrics when evaluating the commit message generation task.} For the task of commit message generation, BLEU and Rouge-L are the most two widely used automation indicators, which are based on the similarity between the generated commit messages and the messages written by developers. Therefore, the quality of the messages written by human developers is the key to whether these indicators are reliable. Unfortunately, according to the results of both our human evaluation and previous research~\cite{nmt}, the quality of commit messages written by developers varies. This leads to the fact that a high metric score does not guarantee that the generated commit message is informative, clear, or contextually appropriate. This finding highlights the need for more robust evaluation metrics that align closely with human judgment in practical use.

\textbf{The quality of human-written commit messages may also affect the performance of models for commit message generation.}
Most existing commit message generation models, especially the ones using machine learning techniques, were trained on datasets consisting of human-written commit messages. If these datasets include poor-quality messages, the models might learn to replicate these inadequacies. This is a critical concern, as the models might generate commit messages that are technically similar to the training data but are not useful in practice. For the task of commit message generation, there is still a lack of a high-quality dataset. Therefore, existing models for commit message generation have certain quality risks, which expect more attention from the researchers.


\section{Limitations}

First, since ChatGPT is a closed-source commercial product, the performance of gpt-3.5-turbo model in commit message generation may be different if the model is executed in multiple rounds. Therefore, inconsistency in the results generated by ChatGPT may affect the reproducibility of the evaluation results due to the non-deterministic nature of LLMs.
Second, this exploratory study only evaluated the zero-shot performance of LLMs. If more strategies are applied to construct diverse prompts, the performance of LLMs in commit message generation might be improved. 
Third, we only selected two LLMs for evaluation. As a preliminary work, we did not evaluate all the existing LLMs as comprehensively as possible. This may lead to specific threats to validity.
Finally, in Phase II of our evaluation, only 366 code difference and their corresponding commit messages out of 7,661 samples were manually evaluated by the first two authors. The evaluation results might be different if more samples are reviewed by these two coders. Human evaluation may introduce subjective bias, and we have adopted some measures to reduce this bias as detailed in the last paragraph of Section \ref{sec:Phase2}.

\section{Conclusions \& Future Work}
\label{sec:future}
Our study has pioneered the exploration of using LLMs for commit message generation. We conducted a 2-phase evaluation on the commit messages generated by LLMs, including metrics evaluation and human evaluation. The experimental results indicate that our research demonstrates the potential of LLMs in automatically generating commit messages. By comparing LLMs, i.e., ChatGPT and Llama 2, against the five baseline methods, we have established their superiority in terms of both quantitative metrics and human preference. This study not only furthers the understanding of LLM applications in practical software development scenarios, but sets a new benchmark for future research in the field.

Furthermore, the results reveal that the quality of human-written commit messages may affect the applicability of the existing metrics for evaluating the generated commit messages and the practical performance of existing models on the commit message generation task. Therefore, for the task of commit message generation, there is still a lack of robust evaluation metrics that align closely with human judgment in practical use.

In the next step, we plan to use more prompt strategies (e.g., CoT~\cite{cot}) to improve the performance of LLMs. Additionally, we intend to develop an LLM-integrated commit message generation method to seamlessly fit into the existing software development life cycle, in order to aid developers in crafting commit messages efficiently and in good quality. 


\section*{Acknowledgments} \label{sec:ack}
This research work is supported by he National Key Research and Development Program of China (No. 2022YFF0902701) and the NSFC (Nos. 62172311 and 62032016).

\balance

\bibliographystyle{IEEEtran}
\bibliography{main}

\end{document}